\documentclass[traditabstract]{aa}  

\usepackage{natbib}
\bibpunct{(}{)}{;}{a}{}{,}    %% natbib cite format used by A&A and ApJ
\usepackage{subfigure}
\usepackage[colorlinks=true, linkcolor=blue, citecolor=blue, urlcolor=blue]{hyperref}
\usepackage{graphicx}
\usepackage{txfonts}
\usepackage{multicol}
\usepackage{multirow}

\begin{document} 

 \title{Impact of seeing and host galaxy into the analysis of
   photo-polarimetric microvariability in blazars
 \thanks{Based on observations collected at the Centro Astron\'omico
    Hispano Alem\'an (CAHA) at Calar Alto, operated jointly by the
    Max-Planck Institut f\"ur Astronomie and the Instituto de
    Astrof\'{\i}sica de Andaluc\'{\i}a (CSIC).}  }
 \subtitle{Case study of the nearby blazars 1ES\,1959+650 and HB89\,2201+044}

 \author{M. S. Sosa\inst{1,2}, C. von Essen\inst{3},
    I. Andruchow \inst{1,2} \and S. A. Cellone\inst{1,2}}
 \institute{Instituto de Astrof\'{\i}sica La Plata, CONICET-UNLP, Paseo del Bosque S/N, B1900FWA La Plata, Argentina\\
              \email{marinasosa@fcaglp.unlp.edu.ar}
        \and Facultad de Ciencias Astron\'omicas y Geof\'isicas, Universidad Nacional de La Plata, Paseo del Bosque S/N, B1900FWA La Plata, Argentina   
        \and Stellar Astrophysics Centre, Department of Physics and Astronomy, Aarhus University, Ny Munkegade 120, DK-8000 Aarhus C, Denmark\\
             }

   \date{Received September 30, 2016 ; accepted July 19, 2017}

\abstract{Blazars, a type of Active Galactic Nuclei, present a
  particular orientation of their jets close to the line of
  sight. Their radiation is thus relativistically beamed, giving rise
  to extreme behaviors, specially strong variability on very short
  time-scales (i.e., microvariability). Here we present simultaneous
  photometric and polarimetric observations of two relatively nearby
  blazars, \mbox{1ES 1959+650} and \mbox{HB89 2201+044}, that were
  obtained using the Calar Alto Faint Object Spectrograph mounted at
  the 2.2~m telescope in Calar Alto, Spain. An outstanding
  characteristic of these two blazars is the presence of well resolved
  host galaxies. This particular feature allows us to produce a study
  of their intrinsic polarization, a measurement of the polarization
  state of the galactic nucleus unaffected by the host galaxy. To
  carry out this work, we computed photometric fluxes from which we
  calculated the degree and orientation of the blazars
  polarization. Then, we analyzed the depolarizing effect introduced
  by the host galaxy with the main goal to recover the intrinsic
  polarization of the galactic nucleus, carefully taking into
  consideration the spurious polarimetric variability introduced by
  changes in seeing along the observing nights. We find that the two
  blazars do not present intra-night photo-polarimetric variability,
  although we do detect a significant inter-night
  variability. Comparing polarimetric values before and after
  accounting for the host galaxies, we observe a significant
  difference in the polarization degree of about 1\% in the case of
  \mbox{1ES 1959+650}, and 0.3\% in the case of \mbox{HB89 2201+044},
  thus evidencing the non-negligible impact introduced by the host
  galaxies. We note that this host galaxy effect depends on the
  weaveband, and varies with changing seeing conditions, so it should
  be particularly considered when studying frequency-dependent
  polarization in blazars.}


\titlerunning{Impact of seeing and host galaxy into the
  microvariability in blazars} \authorrunning{M. S. Sosa et al.}
\maketitle
   
%
%-------------------------------------------------------------------

\section{Introduction}
\label{sec:intro}
Microvariability is defined as the occurrence of rapid changes in the
optical brightness of astrophysical sources with time scales ranging
from some minutes to hours. In particular, this kind of variability
has been observed in blazars
\citep{Miller1989,Romero1999,Andruchow2005}, a subclass of extreme
active galactic nuclei (AGNs) that changes both in optical and
polarized light. High polarization levels ($>3 \%$) and
photo-polarimetric variability are in fact distinctive characteristics
of blazars \citep{Urry1995,Andruchow2003,Cellone2007}. Due to their
favorable orientation, blazars provide a natural laboratory to study
the mechanisms of energy extraction from the central super-massive
black holes and the physical properties of astrophysical jets, also
providing the most adequate testbeds to study their observed
microvariability. For instance, \cite{Romero1995} suggested the
presence of a shocked jet as source of radio photo-polarimetric
variability. Other works proposed that small variations in the
direction of the shocks, that propagate down the relativistic jet,
could produce variations in the observed flux and polarization state
\citep{Marscher1985,Wagner1995,Marscher1996,Andruchow2003}. In both
cases, simultaneous total-flux and polarization microvariability could
be used to confront these models, revealing details about the
fine-structure of the magnetic field in the inner jets.

The spectral energy distribution (SED) of blazars shows two well
defined broad spectral components \citep{Giommi1995}. Depending on the
location of these SED peaks, blazars are classified into low energy
peaked blazars (LBLs) and high energy peaked blazars (HBLs)
\citep{Padovani1995}. While for LBLs the first SED component peaks in
radio to optical and the second component peaks at GeV energies, for
HBLs the first component peaks in UV/X-rays and the second component
peaks at TeV energies. There are, in consequence, some differences
between HBLs and LBLs. Observations of HBLs show statistically lesser
amounts of optical variability and polarized light
\citep{Heidt1996,Heidt1998} than that of LBLs
\citep{Villata2000,Andruchow2005}. \citet{Heidt1998} and
\citet{Romero1999} found that these objects display different duty
cycles and variability amplitudes from those of the LBLs. Such
differences may possibly be attributed to the presence of stronger
magnetic fields in the HBLs \citep{Gaur2012}. Blazars detected so far
at TeV energies are relatively nearby objects, since very high energy
photons are efficiently absorbed by the extragalactic background
light. Due to this closeness, their host galaxies are relatively
bright and spatially resolved; their contribution to the observed
total and polarized flux should (and can) be modelled and
subtracted.

Here we present photo-polarimetric observations of two relatively
nearby \mbox{($z < 0.05$)} blazars, one HBL and one LBL. The main goal
of this work is to characterize their photo-polarimetric behavior,
modelling out the depolarizing effect of their host galaxies, and
considering the contribution to the photo-polarimetric variability of
changing seeing conditions. The HBL blazar \mbox{1ES 1959+650} was
firstly detected at TeV energies by \cite{Aharonian2003}, and has a
redshift of \mbox{$z = 0.048$} \citep{Schachter1993}. The LBL blazar
\mbox{HB89 2201+044} is at \mbox{$z = 0.027$} \citep[][and references
  therein]{Sambruna2007}. It has been classified as a BL object
\citep{Burbidge1987,VeronCetty1989}, and has not yet been detected at
TeV energies. Till date, there are only isolated measurements of the
polarization degree of both blazars, with values of \mbox{$P = 6.9\%$}
for \mbox{1ES 1959+650} \citep{Sorcia2013intro3}, and \mbox{$P = 1.1 -
  1.5\%$} for \mbox{HB89 2201+044} \citep{Brindle1986}. Contrary to
what observations statistically show and models predict, the HBL
should have a low polarization degree, while the LBL should have a
significantly higher value \citep{Urry1995}. Due to their proximity,
both host galaxies have already determined structural parameters, such
as their effective radii and integrated magnitudes
\citep{USO00}. These blazars are relatively nearby objects. In
consequence, two main aspects have to be carefully considered. Firstly,
their host galaxies are bright and have relatively high angular
diameters, potentially introducing a depolarizing effect that can be
translated into an erroneous polarimetric characterization of the
sources, if not properly taken into consideration. This effect is
always present, regardless the observing conditions. Secondly, the
extended surface brightness profile of the host galaxy is relatively
less affected by seeing than the point-like AGN. Therefore, changes in
seeing during the observations could lead to systematic errors in the
photo-polarimetric light curves if not properly accounted for
\citep{CRC00,NPT07,ACR08}. This, in turn, could be wrongly interpreted
as photo-polarimetric variability.

In this work we analyze high-temporal resolution light curves of our
two targets, both at total flux and at polarized flux. This study will
allow us to test and extend the procedure detailed in \citet{ACR08},
to correct for the depolarizing effect of the host galaxy and the
effect introduced by varying seeing conditions, to conduct an adequate
characterization of the polarization states of the two blazars, and to
study their seeing-free microvariability.

We describe the acquired data and instrumental setup, along with our
photometric reduction technique in Section~\ref{sec:OaDR}. In
Section~\ref{sec:Results} we study the photo-polarimetric behavior of
both blazars along the observing campaign, plus the impact of the host
galaxies and seeing on our polarimetric measurements in
Section~\ref{sec:results2}. We end up with a discussion and
conclusions in Section~\ref{sec:Discussion}.
%__________________________________________________________________

\begin{table*}[ht!]
  \caption{List of observed objects during our campaign. From left to
    right the object name, right ascension, declination, redshift,
    visual magnitude, type of source, exposure time, and nature of the
    source. Standard polarimetric stars were obtained from
    \cite{SEL92} and \cite{TBW90}.}
  \label{object-data}
  \centering 
  \scalebox{0.9}{
  \begin{tabular}{l c c c c c l l} 
    \hline\hline
Name  & $\alpha_{2000.0}$ & $\delta_{2000.0}$ & $z$ & $m$ & Type  & Exposure time & nature of the source \\
      & (h min s)         & ($^\circ ~~ ' ~~ ''$) & & (mag) &   &~~~~ (s)~~~~ & \\
 \hline 
 1ES 1959+650  & $19~59~35.00$  & $+65~00~14.0$ & $0.048$ & $15.38$ (R)     & HBL   & 60-120       & science target \\
 HB89 2201+044 & $22~04~17.65$  & $+04~40~02.0$ & $0.027$ & $17.18$ (R)    & LBL   & 60-120       & science target \\
 BD+59389        & $02~02~42.06$  & $+60~15~26.5$ & -       & $10.34$ (V) & LSS   & $0.2-4$      & polarized star \\
 HD+204827       & $21~28~57.70$  & $+58~44~24.0$ & -       & $7.93$ (V)  & LSS   &   $0.2-4$    & polarized star  \\
 HD+212311       & $22~21~58.60$  & $+56~31~53.0$ & -       & $8.10$ (V)  & LSS   & $0.2-4$      & unpolarized star  \\
 \hline
  \end{tabular}
}
\end{table*}

\section{Observations and data reduction}
\label{sec:OaDR}

\subsection{Targets and observing strategy}
\label{sec:tar_obs_strat}

\begin{figure}[ht!]
  \centering
  \includegraphics[width=.438\textwidth]{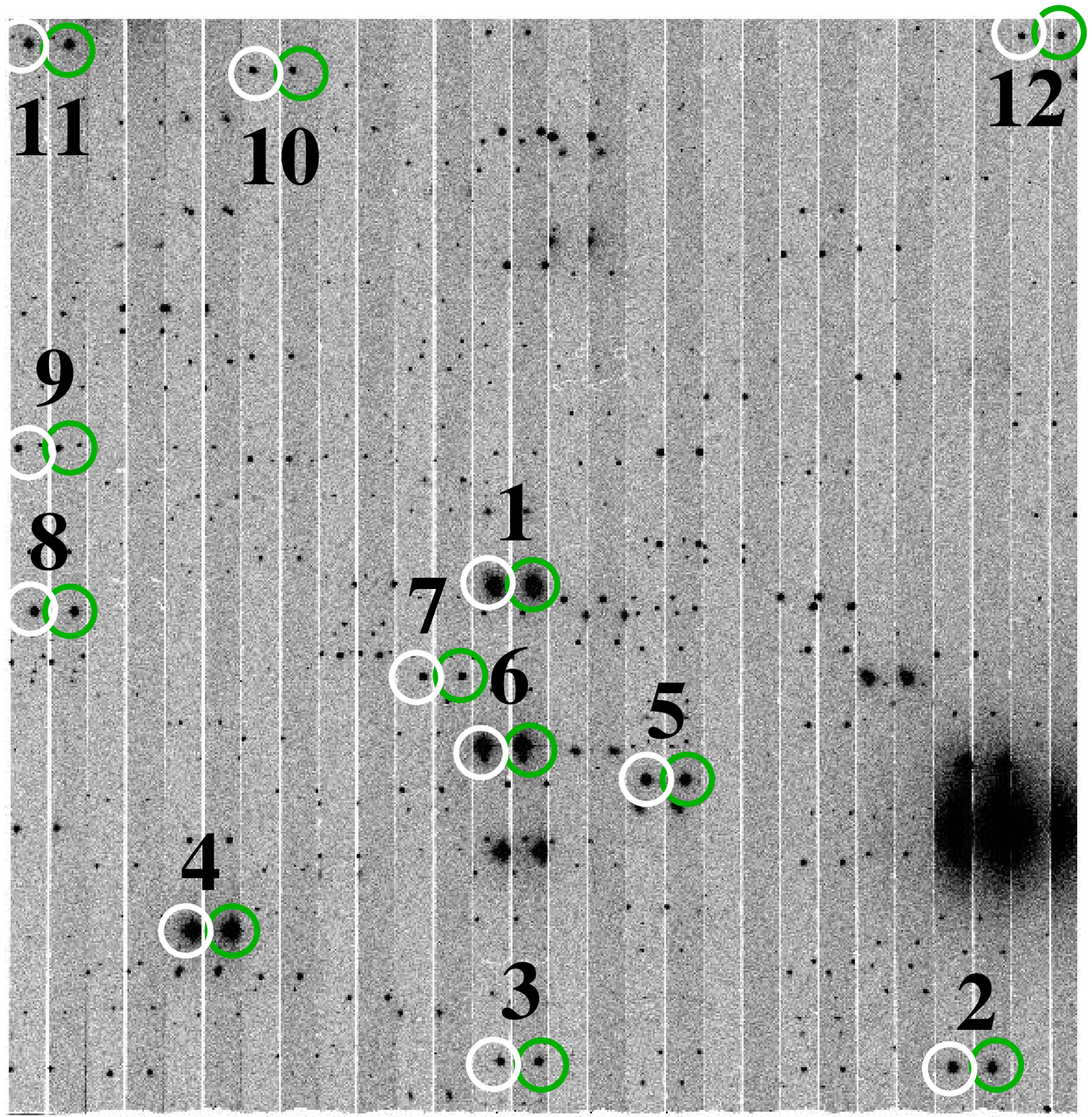}
  \includegraphics[width=.43\textwidth]{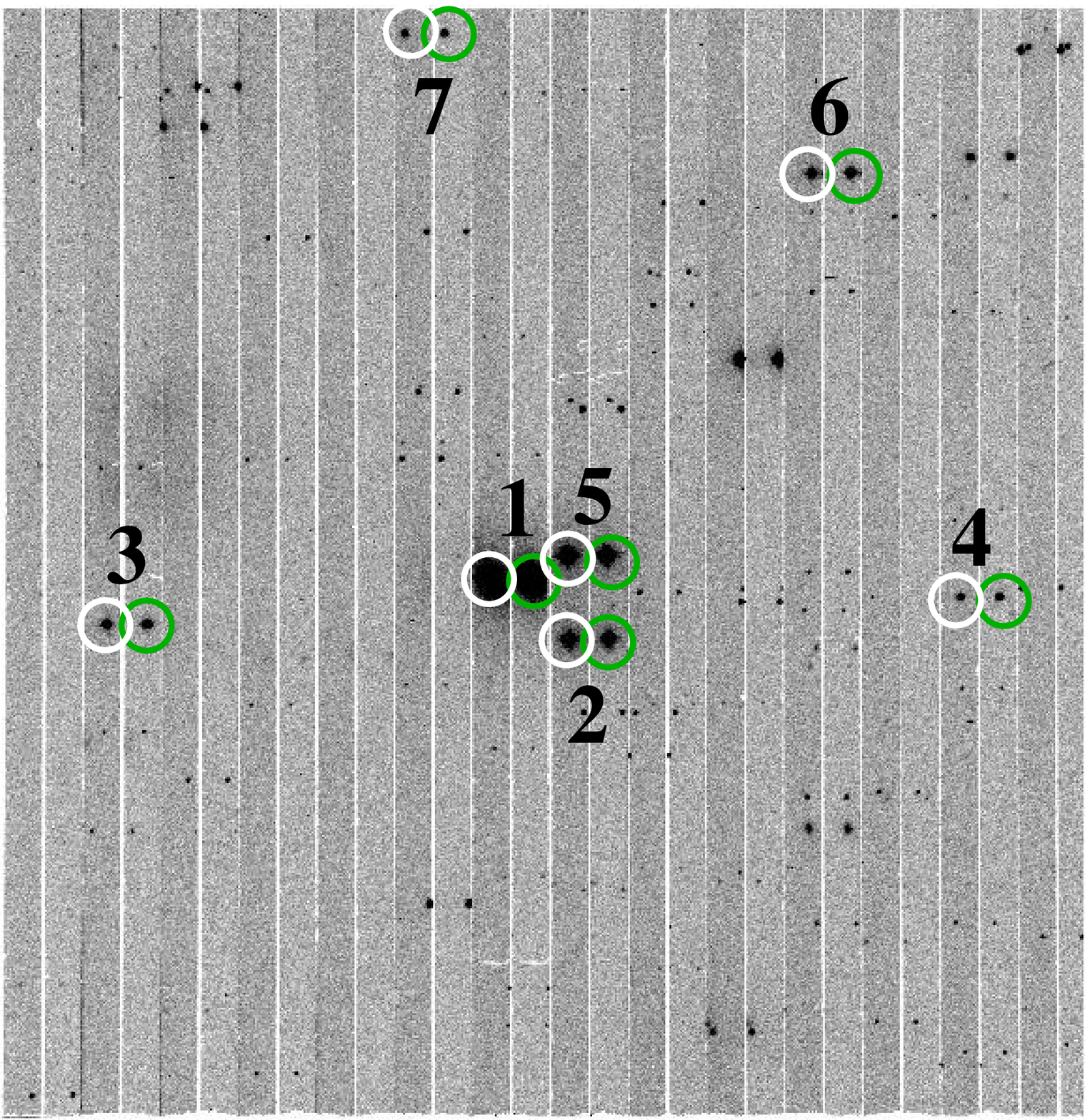}
  \caption{\label{fig:fov} The polarimetric frames of the two fields
    for \mbox{1ES 1959+650} {\it (top)}, and \mbox{HB89 2201+044} {\it
      (bottom)}, observed in the $R$ band. In both cases, the white
    and green circles indicate the locations of the blazar and field
    stars for the ordinary and extraordinary images, respectively,
    prior to the virtual masking. Labeled with 1 are the blazars, and
    with numbers larger than 1 the field stars whose fluxes were
    measured. The field of view is $9$ $\times$ $9$ arcmin, and East
    is up and North is to the right.}
\end{figure}

We observed the blazars \mbox{1ES 1959+650} and \mbox{HB89 2201+044}
for six consecutive dark nights between July 29 and August 3$^{rd}$,
2011. We carried out our observations using the Calar Alto Faint
Object Spectrograph (CAFOS) mounted at the Calar Alto 2.2~m telescope
in its imaging and polarimetric modes \citep{Meisenheimer1998}. CAFOS
has a Wollaston prism plus a rotatable half-wave plate \citep{PT2011},
that produces two orthogonal polarized images, henceforth ordinary (O)
and extraordinary (E) beams, of each object on the focal plane. This
allowed us to simultaneously record photometric and polarimetric
data. The detector used was the 2SITE\#1d charge-coupled device (CCD),
with a total size of \mbox{2k$\times$2k} pixels, each one with a size
of \mbox{24 micron}, from which only the central \mbox{1k$\times$1k}
was read (see Section~\ref{sec:polarimetry} for further details on
this choice). The CCD has a gain of \mbox{2.3 e-/ADU} and a readout
noise of \mbox{5.06 e-}. To avoid any overlap of the O/E images we
placed a physical mask with alternate blind and clear stripes of about
\mbox{20 arcsec} width each. Although the observing technique implies
a lost of half of the field of view, it significantly improves the
final signal-to-noise (S/N) ratio of the data. Figure~\ref{fig:fov}
shows the field of views of the blazars: \mbox{1ES 1959+650} {\it
  (top)} and \mbox{HB89 2201+044} {\it (bottom)},
respectively. Table~\ref{object-data} summarizes the most relevant
parameters of all the objects observed during the campaign.

Multicolor polarimetry can provide information about the innermost
part of an AGN. Thus, the wavelength dependence of polarization of
blazars has been studied by several authors \citep[see e.g.,][and
  references therein]{Brindle1986}. Among others, it can reveal the
presence of more than one synchrotron component, and the dilution by
thermal radiation from the parent galaxy \citep{Maza1978}. In
consequence, we observed using $R$ and $B$ filters. Exposure times
ranged between 60 and 120 seconds, strongly depending on the
atmospheric conditions, the altitude of the blazars during the
observations, and the chosen filter. We acquired sky flats and bias
frames on regular basis. Atmospheric conditions were different
throughout the observing time, from fairly good (no clouds, seeing
full width at half maximum, \mbox{FWHM $\sim$ 1} arcsec) to rather
poor (cirrus, \mbox{FWHM $\ga$ 2.5} arcsec). During poor observing
conditions, we observed only in the $R$ band, where atmospheric
extinction and scattering of spurious light by dust particles in the
atmosphere are less dominant.

For the image pre-processing we used standard IRAF routines. We
corrected all science frames by bias and flats in the usual
way. However, for the data extraction and analysis we used IRAF tasks
created by our group. Previous to the photometry, it was necessary to
multiply the images by a virtual mask. This consists of an image with
alternate stripes of $\sim 20$ arcsec width, with pixel values set to
either one or zero. Our IRAF routines are tailored to appropriately
handle zero-valued pixels within the aperture and the sky annulus. If
there is one or more zero-valued pixels within the photometric
aperture, the magnitude is not defined and the result from that
aperture is discarded. Nonetheless, the final aperture size was well
contained within the stripes, and magnitudes computed from apertures
up to 10 arcseconds ($\sim$5 FWHM) were always defined. This procedure
avoids contamination of the E images, in the aperture and sky annulus
used for photometry by the O images, and vice-versa. Afterwards, we
obtained the instrumental magnitudes and polarimetric data
corresponding to the E and O images for the two blazars on each
frame. The same procedure was carried out on stars evenly distributed
on the field and suitably placed within the mask stripes (i.e. we
rejected stars close to the edge of the stripes). We used them as
estimators of the instrumental and foreground polarization, and to
produce the differential light curves in both bands along the whole
campaign.

To maximize the S/N ratio of our measurements, we carried out a careful
selection of the aperture radius. We measured photometric and
polarimetric quantities integrating blazar and stellar fluxes within
15 apertures, starting from \mbox{1 arcsec} up to \mbox{10 arcsec}. To
select the final aperture we followed \cite{H89} and our own error
analysis. While the former requires an aperture radius close to the
FWHM of the sources, the latter requires the minimization of the error
in the polarization degree and the polarization angle. Although the
aperture that minimized these quantities slightly changed along the
nights (just because the photometric conditions during the
observations changed), the optimal aperture radius turned out to be
always around \mbox{3 arcsec}, which corresponds to the largest seeing
value along the whole campaign. Since our goal is to analyze
polarimetric variability and compare this along the observing nights,
to carry out a photo-polarimetric analysis as homogeneous as possible
we fixed the aperture radius to \mbox{3 arcsec} in both data sets and
throughout the campaign.

\subsection{Stokes parameters}

To obtain the normalized Stokes parameters we used four frames, each
with a different position angle of the rotating plate (0, 22.5, 45,
and 67.5 degrees). Their mathematical expressions are:
\begin{equation}
\begin{split}
& Q=\frac{R_{\mathrm{Q}}-1}{R_{\mathrm{Q}}+1}\,,
& U=\frac{R_{\mathrm{U}}-1}{R_{\mathrm{U}}+1}\,,\\
\end{split}
\end{equation}
\noindent where
\begin{equation}
\begin{split}
& R_{\mathrm{Q}}^2=\frac{I_{0}^{O}/I_{0}^{E}}{I_{45}^{O}/I_{45}^{E}}\,,
& R_{\mathrm{U}}^2=\frac{I_{22.5}^{O}/I_{22.5}^{E}}{I_{67.5}^{O}/I_{67.5}^{E}}\,,\\
\end{split}
\end{equation}
\noindent $I_{\beta}^{O}$ y $I_{\beta}^{E}$ are the object ordinary
and extraordinary integrated fluxes, respectively, and $\beta$ is the
position angle of the half-wave plate
\citep{Zapatero-2005,Andruchow2011}. Based on these parameters, we
calculated the degree of polarization and corresponding position angle
in the usual way:
\begin{equation}
\begin{split}
& P=\sqrt{Q^2+U^2}\,,\\
& \Theta=\frac{1}{2}\mathrm{arctan}\left(\frac{U}{Q}\right) \,.\\
\end{split}
\end{equation} 
\noindent Error estimates for both parameters were computed using
standard error propagation techniques. However, our final expressions
were verified with and compared to the ones available in
\cite{PR2006}. This procedure was carried out equivalently for the two
blazars, all the field stars labeled in Figure~\ref{fig:fov}, and the
polarized and unpolarized standard stars. To analyze the consistency
of our error determination, we checked (and also always verified) that
the magnitude of the individual polarimetric errors was comparable to
the magnitude of the standard deviation of the polarimetric points
within a given night.

\subsection{Extrinsic polarization: Instrumental and foreground polarization}

For the calibration and transformation of the data to the standard
system \citep[N-E-S-W,][]{Lamy1999}, we observed highly polarized and
unpolarized standard stars (Table~\ref{object-data}) cataloged under
\cite{SEL92} and \cite{TBW90}. Following \cite{PR2006} and
\cite{PT2011}, we analyzed the data obtained for the unpolarized
standard stars to quantify the instrumental polarization per
filter. We observed a standard unpolarized star every night, that was
placed at exactly the same location than the blazars, coincidentally
being the center of the CCD. CAFOS is known to have a dependency of
the instrumental polarization with the position over the CCD,
increasing towards the edges \citep[see, e.g.,][for a careful
  characterization of the instrumental polarization of
  CAFOS]{PT2011}. Thus, our observing strategy was diagramed to
properly characterize the instrumental polarization at the position of
the target. A characterization of the instrumental polarization of
CAFOS is beyond the scope of this work. From these observations we
calculated the Stokes parameters in the usual way per night and per
filter, and we averaged them along the whole campaign. In agreement
with previously reported values, we found a low contribution of the
instrumental polarization at the center of the CCD. For the $R$ band
we found \mbox{Q$_{\mathrm{instr}}$ = 0.029\%} and
\mbox{U$_{\mathrm{instr}}$ = -0.015\%}, and for the $B$ band we found
\mbox{Q$_{\mathrm{instr}}$ = -0.14\%} and \mbox{U$_{\mathrm{instr}}$ =
  -0.013\%}. As a complementary checkup of the expected polarization
behavior of CAFOS, we computed the polarization degree of all the
field stars. Assuming that these are unpolarized, we see a clear
dependency of their polarization degree with their distance to the
center of the CCD.

We also estimated a lower limit to the foreground polarization using
the stars $5$, $6$ and $7$ in the case of \mbox{1ES 1959+650}, and
stars $2$ and $5$ in the case of \mbox{HB89 2201+044}. Labels are as
in Figure~\ref{fig:fov}. Foreground polarization is small:
\mbox{$0.7\%$} in $R$ and \mbox{$0.9\%$} in $B$ for \mbox{1ES
  1959+650} and and \mbox{$0.3\%$} in $R$ and \mbox{$0.4\%$} in $B$
for \mbox{HB89 2201+044}. This is consistent with the upper limits
estimated using $P_{\mathrm{max}} \leq 9 E_{\mathrm{B-V}}$
\citep{Ho96,serk1975}, with the $E_{\mathrm{B-V}}$ indexes obtained
from the NASA/IPAC Extragalactic Database \citep{Schlafly2011}, giving
\mbox{$P_{\mathrm{V}} < 1.4\%$} for \mbox{1ES 1959+650} and
\mbox{$P_{\mathrm{V}} <0.34\%$} for \mbox{HB89 2201+044}. Thus, from
now on all the polarimetric quantities are corrected by instrumental
and foreground polarization in the following fashion:
\begin{equation}
\begin{split}
& Q_{\mathrm{corr}} = Q_{\mathrm{obs}} - Q_{\mathrm{instrumental}} - Q_{\mathrm{foreground}}\,,\\
& U_{\mathrm{corr}} = U_{\mathrm{obs}} - U_{\mathrm{instrumental}} - U_{\mathrm{foreground}}\,.\\
\end{split}
\end{equation} 
\noindent In all cases, position angles were transformed to the
Standard system using data from highly polarized standard stars.

Finally, we estimated the unbiased degree of polarization, $P_{0}$. We
use the expression,

\begin{equation}
P_0 = \sqrt{P^2 - a * P_{\mathrm{err}}^2}~,
\end{equation}
as specified in \cite{Simmons1985}. $P_{0}$ was computed using the
maximum likelihood estimator that can be found in their work
($a=1$). For both blazars, the correction is one order of magnitude
smaller than the polarimetric errors. Thus, we did not apply the
correction since it is statistically negligible. In the case of
\mbox{HB89 2201+044}, we did not calculate this correction in the
B-band because there are not enough data points.

%________________________________________________________________

\section{Photometric and polarimetric analysis}
\label{sec:Results}

\subsection{Photometry}
\label{sec:photom}

\begin{figure*}[ht!]
  \centering
  \includegraphics[width=.475\textwidth]{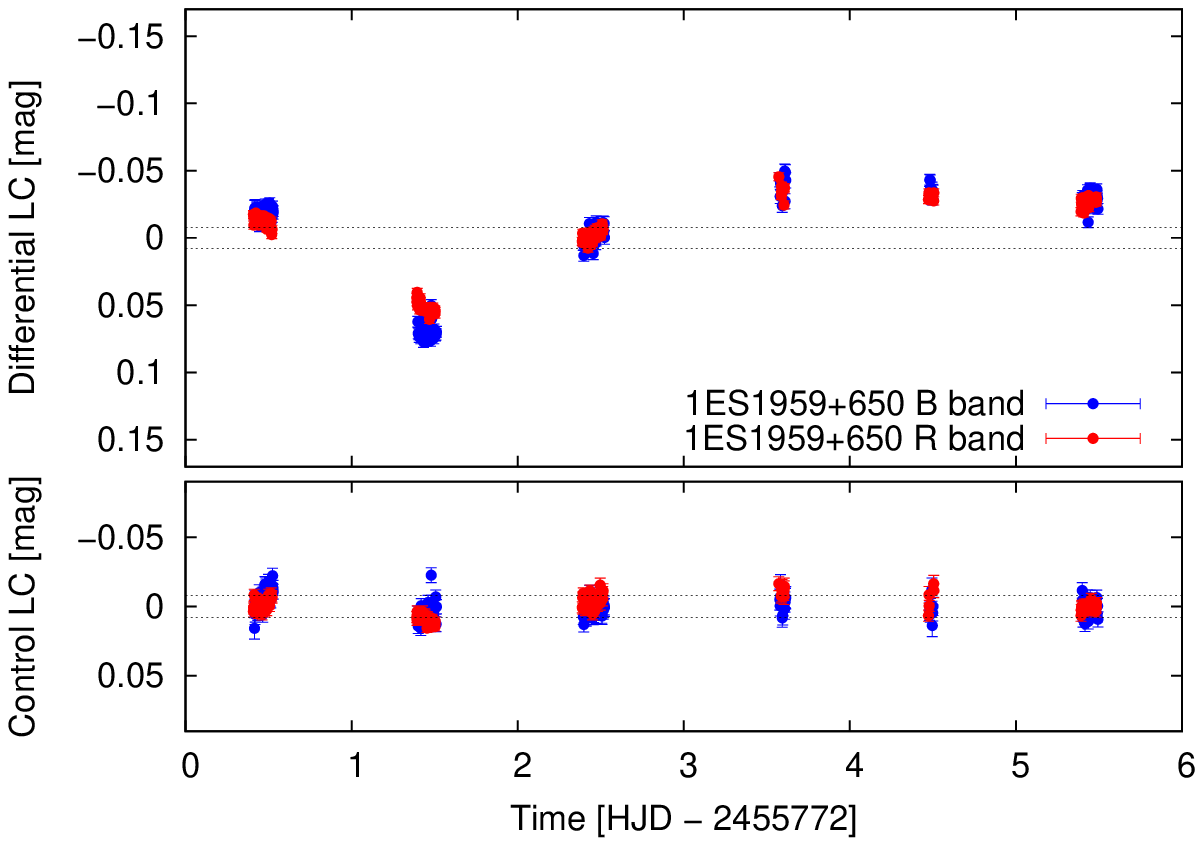}%
  \includegraphics[width=.475\textwidth]{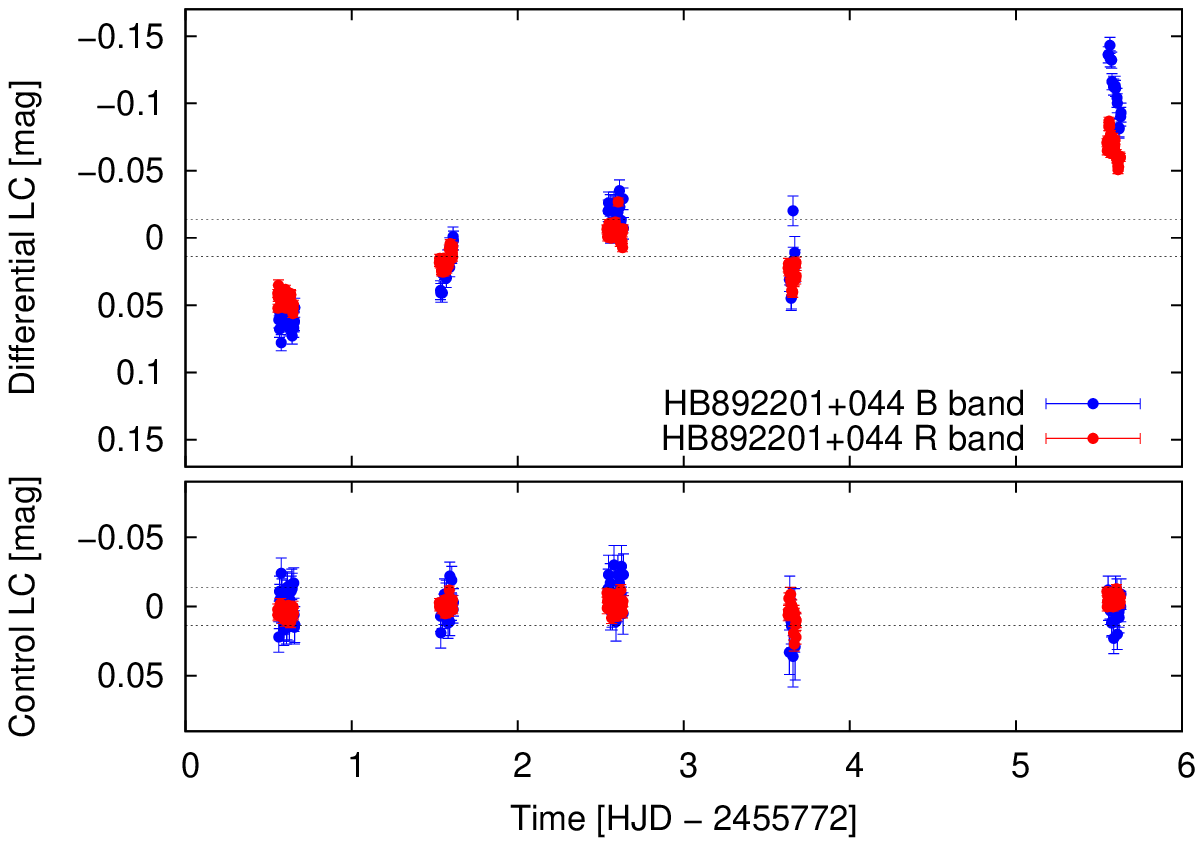}
  \caption{\label{fig:FD} Differential (top sub-panels) and control
    (bottom small sub-panels) light curves (LC) in magnitudes as a
    function of the Heliocentric Julian Date (HJD) for the two blazars
    observed in this work. {\it Left:} \mbox{1ES 1959+650}, $B$ and
    $R$ bands. {\it Right:} \mbox{HB89 2201+044}, $B$ and $R$
    bands. Horizontal black-dashed lines show two times the $B$
    standard deviation of the control light curves. Red data points
    correspond to $R$ band data, while blue data points to the $B$.}
\end{figure*}

The differential light curves for the $B$ and $R$ filters for both
blazars, spanning the whole observational campaign, are shown in
Figure~\ref{fig:FD} \citep[see][for details about the construction of
  the light curves]{Andruchow2011}. Particularly, to construct the
differential and control light curves of \mbox{1ES 1959+650}, we used
the stars labeled 2 and 8, respectively (see Figure~\ref{fig:fov}) for
both $R$ and $B$ bands. For \mbox{HB89 2201+044}, the comparison star
is number 3, while the control star is number 4. This applies to both
bands.

To characterize the variability in our photo-polarimetric data we used
the scaled $C$ criterion \citep{HWM88}, one of the most reliable tools
for this kind of analysis \citep{Zibecchi2017}. The criterion is
defined as the quotient between the standard deviation of the
differential light curve of the blazar (DLC), $\sigma_\mathrm{DLC}$,
and the standard deviation of the control light curve (CLC),
$\sigma_\mathrm{CLC}$, scaled by a factor $\Gamma$ that takes into
consideration the different relative brightnesses between the AGN and
the comparison and control stars. This is represented as the scaled
confidence parameter, $\mathrm{C}_\Gamma$. When $\mathrm{C}_{\Gamma}
\ge 2.576$, variability is detected at least with a \mbox{99.5\%}
confidence level.

The standard deviations of the final control light curves are
\mbox{$\sigma$ = 7.9} milli-magnitudes (mmag) in the $B$ band and
\mbox{$\sigma$ = 6.9 mmag} in the $R$ band for \mbox{1ES 1959+650},
and \mbox{$\sigma$ = 13.6 mmag} in $B$ and \mbox{$\sigma$ = 6.0 mmag}
in $R$ for \mbox{HB89 2201+044}.

The behavior of the $B$ and $R$ bands is similar. In consequence,
Table~\ref{tabla-1959_2201} shows the results for \mbox{1ES 1959+650}
and \mbox{HB89 2201+044} for the $R$ band alone. In both bands,
neither of the two blazars presented intra-night variability. However,
in the case of \mbox{1ES 1959+650} there is evidence of variability on
longer time scales, manifested as a sustained decrease and increase of
flux \mbox{($\Delta R,B \approx 0.1$ mag)} that took place during the
first 3 nights. For \mbox{HB89 2201+044}, during the last 2 nights we
detected an increase of flux \mbox{($\Delta R,B \approx 0.15$
  mag)}. In the case of \mbox{1ES 1959+650}, when only one individual
night is evaluated no variability is observed, while analyzing the
whole campaign returns $\mathrm{C}_{\Gamma} = 2.01$. The corresponding
value for \mbox{HB89 2201+044} is $\mathrm{C}_{\Gamma} = 9.580$. Thus,
both blazars show inter-night variability, although for \mbox{1ES
  1959+650} it might be marginal. In the particular case of \mbox{HB89
  2201+044}, contrary to \mbox{1ES 1959+650} we detected intra-night
variability in two of the six nights. At this stage it was unclear to
us if this observed variability was produced by physical processes
occurring in the source, or were the result of fluctuations in the
seeing during the observing nights. This point is addressed later.

\begin{table}[ht!]
  \caption{Variability parameters for the $R$ light curves on
    different nights for \mbox{1ES 1959+650} {\it (top)} and
    \mbox{HB89 2201+044} {\it (bottom)}. From left to right we
    provide date (Col. 1), target-comparison light curve dispersion
    (Col. 2), control-comparison light curve dispersion (Col. 3),
    scaled confidence parameter (Col. 4), variability classification
    following adopted criterion (Col. 5) and number of $R$ band data
    points (Col. 6). MARG. corresponds to marginal detection of
    variability.}
  \label{tabla-1959_2201}
  \centering 
 % \small{
\begin{tabular}{c c c c c c l l}
\hline\hline
Date   & $\sigma_\mathrm{DLC}$ &  $\sigma_\mathrm{CLC}$ & $\mathrm{C}_{\Gamma}$ & Variable?  & N \\
(mm/dd/yyyy)&  ($R$)             & ($R$)           &                        &            &   \\
\hline
\multicolumn{6}{c}{\mbox{1ES 1959+650}}\\
\hline
07/29/2011 & 0.004 & 0.004 & 0.986 & NO & 33\\
07/30/2011 & 0.004 & 0.003 & 1.263 & NO & 32\\
07/31/2011 & 0.004 & 0.005 & 0.831 & NO & 36\\
08/01/2011 & 0.007 & 0.017 & 0.436 & NO & 12\\
08/02/2011 & 0.005 & 0.033 & 0.164 & NO & 12\\ 
08/03/2011 & 0.003 & 0.002 & 1.514 & NO & 28\\
\\
WC & 0.031 & 0.016 & 2.010 & MARG. & 153\\
\hline
\multicolumn{6}{c}{\mbox{HB89 2201+044}}\\
\hline
07/29/2011 & 0.005 & 0.005 & 1.170 & NO  & 27\\
07/30/2011 & 0.006 & 0.002 & 2.951 & YES(?) & 32\\
07/31/2011 & 0.005 & 0.005 & 2.035 & NO  & 36\\
08/01/2011 & 0.007 & 0.007 & 1.203 & NO  & 20\\
08/03/2011 & 0.009 & 0.003 & 4.478 & YES(?) & 32\\
\\
WC & 0.039 & 0.005 & 9.580 & YES & 147\\
\hline
  \end{tabular}
\end{table}

\subsection{Polarimetry}
\label{sec:polarimetry}

\begin{figure*}[ht!]
  \centering
  \subfigure[]{\includegraphics[width=.45\textwidth]{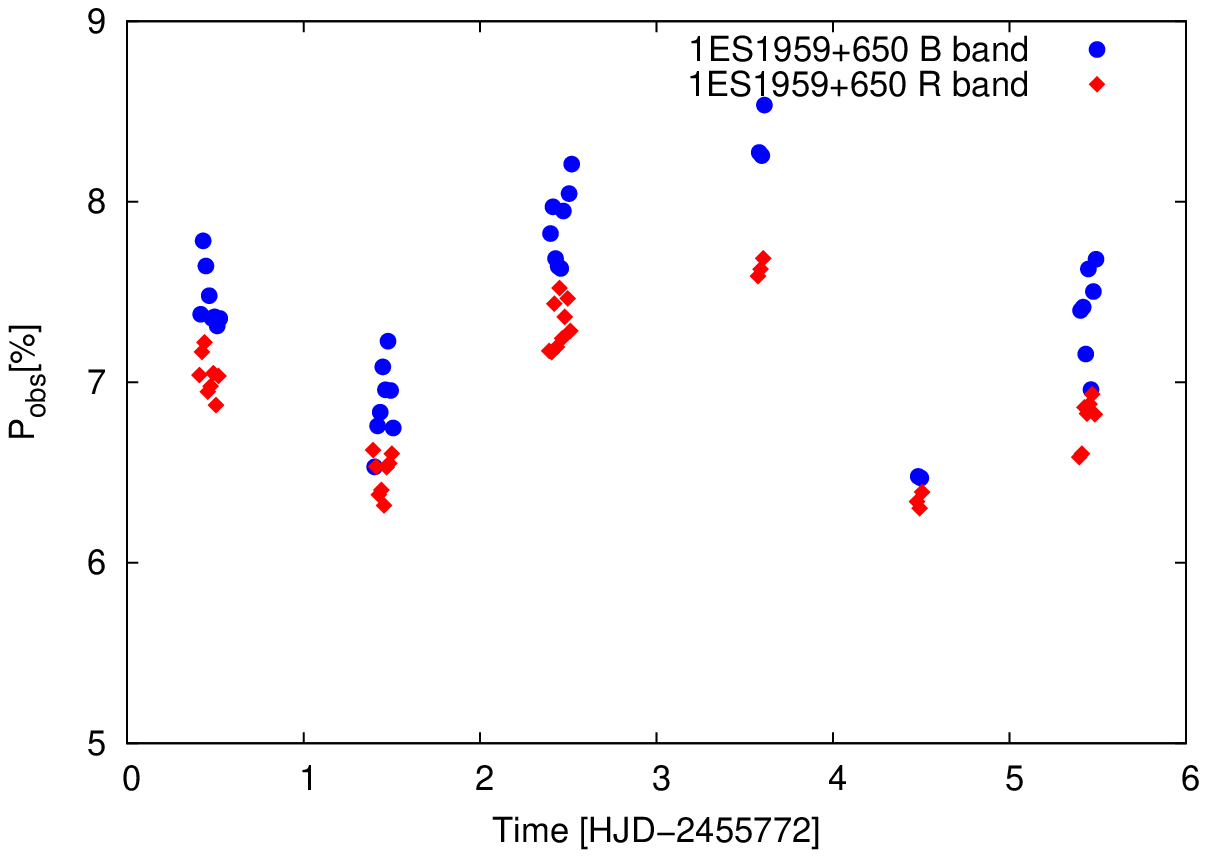}}
  \subfigure[]{\includegraphics[width=.45\textwidth]{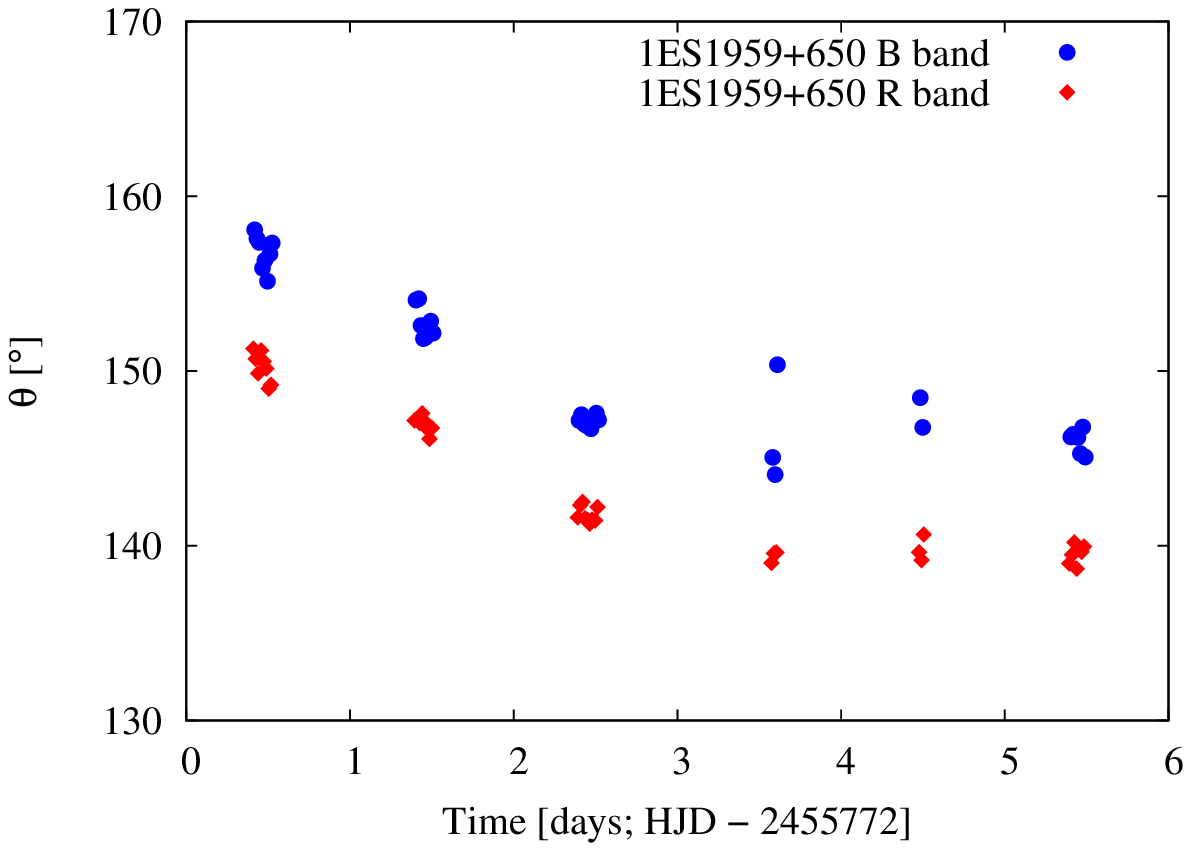}}
  \subfigure[]{\includegraphics[width=.45\textwidth]{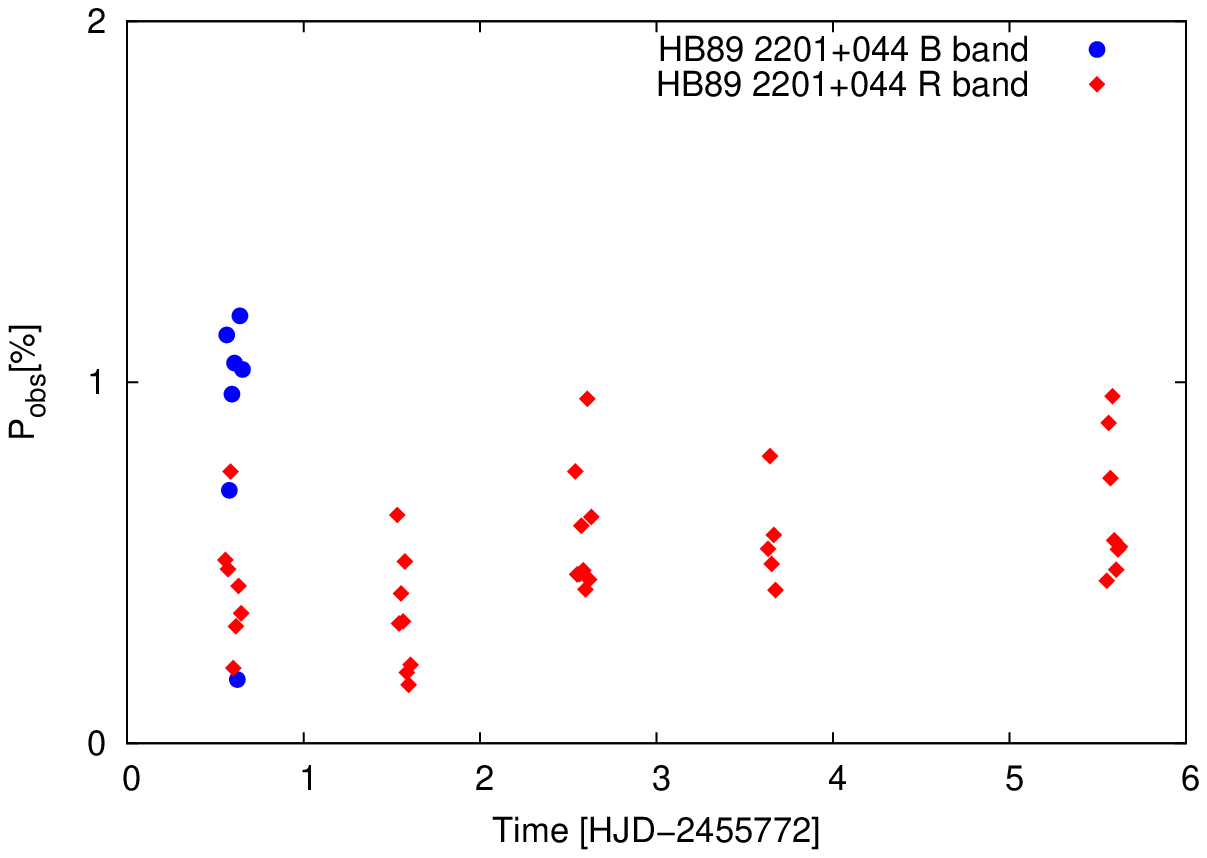}}
  \subfigure[]{\includegraphics[width=.45\textwidth]{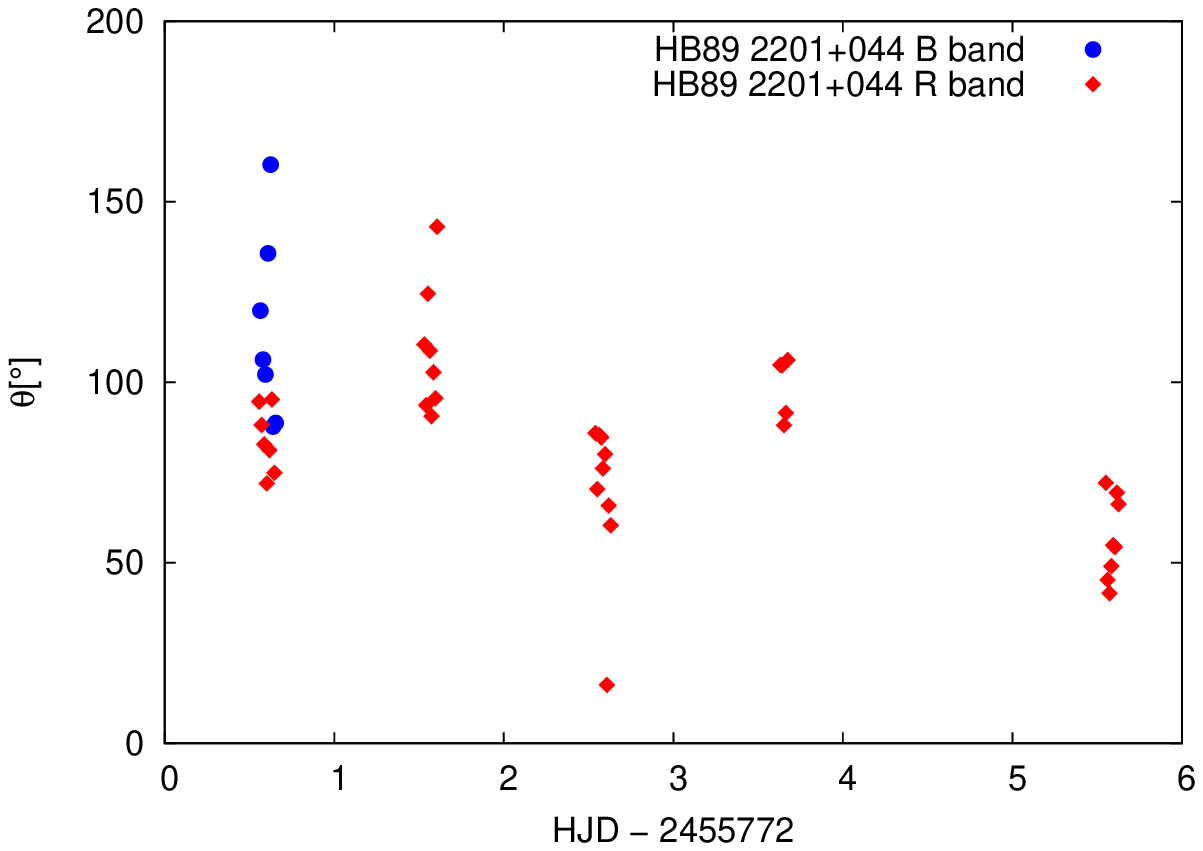}}
  \caption{\label{fig:pol}Time evolution of the polarimetric
    parameters for the $B$ (blue circles) and $R$ (red diamonds)
    bands, corrected by instrumental and foreground polarization. {\it
      Top:} the polarization degree in percentage, $P$ [\%] {\it
      (left)} and the polarization angle in degrees, $\Theta$
    [$\circ$] {\it (right)} for \mbox{1ES 1959+650}. {\it Bottom:}
    equivalently, but for \mbox{HB89 2201+044}.}
\end{figure*}

To characterize the polarization state of the blazars, as specified in
Sect.~\ref{sec:OaDR} we calculated their linear polarization, $P$, and
position angle, $\Theta$. The effect of the galaxy is more evident in
$P$ \citep{ACR08,Cellone2007}. In consequence, along this work we will
specify quantities in ($P$,$\Theta$) rather than in ($Q$,$U$). Due to
poor observing conditions we only have polarimetric data along the
whole campaign in the $R$ band for \mbox{1ES 1959+650}. In the case of
\mbox{HB89 2201+044}, on the fifth night we lack polarimetric
data. Their mean values and standard deviations are summarized in
Table~\ref{tab:polres}, and their temporal evolution can be seen in
Figure~\ref{fig:pol}.

\begin{table}[ht!]
  \caption{Mean values for the polarization degree, $P$, and polarization
    angle, $\Theta$, for the blazars \mbox{1ES 1959+650} and
    \mbox{HB89 2201+044}.}
  \centering
  \begin{tabular}{c c c c}
    \hline \hline
    Blazar                 &     $\langle P \rangle$      &    $\langle \Theta \rangle$  & Band  \\
    \hline
    \mbox{1ES 1959+650}  & 6.97 $\pm$ 0.50 & 145.40 $\pm$ 4.66 & $B$ \\
    \mbox{1ES 1959+650}  & 6.17 $\pm$ 0.41 & 144.33$\pm$ 4.75 & $R$ \\
    \mbox{HB89 2201+044} & 0.70 $\pm$ 0.46 & 168.52 $\pm$ 32.28 & $B$ \\
    \mbox{HB89 2201+044} & 0.38 $\pm$ 0.30 &  188.44 $\pm$ 37.45 & $R$ \\
    \hline
  \end{tabular}
  \label{tab:polres}
\end{table}

We analyzed the behavior of $P$ and $\Theta$ throughout the whole
campaign. For \mbox{1ES 1959+650}, $P$ is relatively high \citep[$\sim
  7\%$, see, e.g.][for comparable results]{Andruchow2005}. A visual
inspection of the polarimetric data along the campaign shows some
inter-night variability, while the angle presents a slight
rotation. On the contrary, \mbox{HB89 2201+044} seems to be steady and
shows an erratic behavior for the polarization angle with large error
bars, which can be attributed to the fact that this parameter is not
well defined because of the low polarization.

To estimate if both blazars show inter and intra-night polarimetric
variability, we carried out a statistical analysis fully described in
\cite{Kesteven1976}. In this case, a given source is qualified as
variable if the probability of exceeding their $\chi^2$ value is
smaller than $0.1\%$, and not variable if the probability is larger
than $0.5\%$. Due to the scarcity of polarimetric data for the $B$
band in \mbox{HB89 2201+044}, we carried out this analysis exclusively
in the $R$ band. The analysis of intra-night variability retrieved
probability values between \mbox{$18\%$} and \mbox{$86\%$} for
\mbox{1ES 1959+650}, and between \mbox{$30\%$} and \mbox{$76\%$} for
\mbox{HB89 2201+044}, clearly favoring the absence of intra-night
variability. When the whole campaign is analyzed, probability values
of 10$^{-8}$\% and 10$^{-45}$\% are obtained for both blazars,
favouring the presence of inter-night variability.

\subsubsection{Impact on $P$ of the aperture choice}

\begin{figure}[ht!]
  \centering
  \includegraphics[width=.45\textwidth]{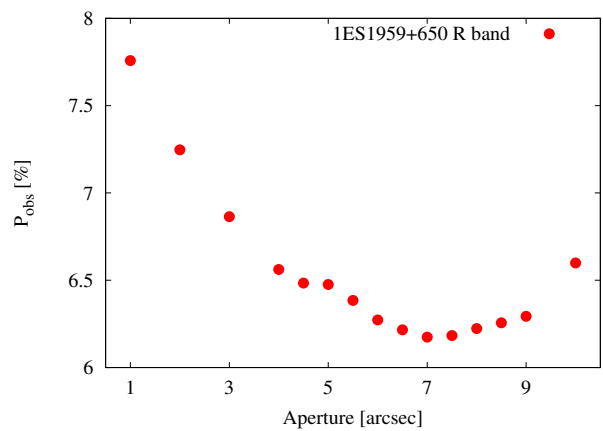}
  \caption{\label{fig:PAbert}Behavior of the polarization degree with
    the aperture for the blazar \mbox{1ES 1959+650} in the $R$
    band.}
\end{figure}

While analyzing the depolarizing effect that is introduced by the host
galaxy on the AGN, careful considerations have to be taken into
account regarding seeing and aperture. Seeing affects in a different
way the AGN (point source) and the galaxy (extended
source). Consequently, a variation in seeing introduces (or removes) a
percentage of unpolarized light within the aperture from the galaxy
which in general will be different to the introduced (or removed)
percentage of light from the nucleus \citep[see][for the impact of
  seeing in this type of measurements]{ACR08}. Therefore, to quantify
how much the host galaxy affects our derived values of the
polarization state of both blazars, we analyzed only the night that
showed almost negligible seeing variability (August 3$^{rd}$,
2011). In this way, the variability to be measured is most likely due
to the intrinsic changes in the polarization state of the blazars and
not due to changes caused by our Earth's atmosphere. As it is
expected, we observed a decrease of the degree of polarization when
larger apertures are being used, revealing the impact of the host
galaxies in our measured values. This effect can be appreciated in
Figure~\ref{fig:PAbert} in the case of \mbox{1ES 1959+650} for the $R$
band, and its detection is independent of the choice of filter or
blazar. Altogether, these results show that one has to choose a
reliable criterion before fixing the aperture to extract fluxes from
photo-polarimetric data. Our particular choice has been already
specified in Sect.~\ref{sec:tar_obs_strat}. It is worth to mention
that $\Theta$ is not affected by changes in the photometric aperture.

\section{Results}
\label{sec:results2}

As previously mentioned, two (related) effects should be considered
when dealing with polarimetric time series of AGN with prominent host
galaxies: the host starlight introduces a depolarazing effect, and its
amount will be variable if seeing changes along the
observations. Moreover, both the depolarization and its variations due
to seeing will depend on wavelength. Similar effects apply to
photometric light-curves, where the host starlight dilutes any
intrinsic AGN variability, and introduces spurious variability under
seeing fluctuations. In this Sect. we will carefully analyze these
effects.

\subsection{Determination of observed structural quantities}
\label{sec:params_struc}

To recover the intrinsic polarization of the AGNs we determined the
structural parameters of both blazar host galaxies. To this end we
combined the images per blazar and per filter that were taken without
the polarizer (2 images in $B$ band and 2 images in $R$ band for
\mbox{1ES 1959+650}, and 6 images in $B$ and 2 images in $R$ for
\mbox{HB89 2201+044}). These images were taken during the night that
presented not only the lowest seeing but also the lowest seeing
variability (August 3$^{rd}$, 2011). In the particular case of
\mbox{HB89 2201+044}, eastwards from the blazar a bright star can be
observed. We considered this, and the effect that this introduces over
the determination of the intrinsic parameters of the host galaxy, as
addressed in the next Sect.

To determine the structural parameters of the host galaxies we used
the combined power of IRAF's tasks {\it ellipse} and {\it
  nfit1d}. Before carrying out the isophotal fit we removed the field
stars and the neighboring galaxies with a mask. During this process we
chose appropriate functions that best match the combined surface
brightness profile of the host galaxy, and the surface brightness
profile of the AGN, combined with the changes in brightness induced by
our Earth's atmosphere.

For the AGN we considered a Gaussian function with the following
parameters:
\begin{equation}
g(a_{\mathrm{AGN}})=g_{\mathrm{0}}\times\mathrm{e}^{-(a_{\mathrm{AGN}}/2\sigma)^2},
\end{equation}
\noindent where $a_{\mathrm{AGN}}$ corresponds to the semi-major axis
of the AGN, \mbox{FWHM = 2$\sqrt{2\,\ln(2)}~\sigma$} accounts for the
seeing (FWHM), and $g_{\mathrm{0}}$ corresponds to the amplitude of
the Gaussian. In this case, for \mbox{1ES 1959+650} we considered
\mbox{FWHM = 1.87 pixels} and \mbox{FWHM = 2.53 pixels} for the $R$
and $B$ bands, respectively, and for \mbox{HB89 2201+044} the
derived values were \mbox{FWHM = 2.18 pixels} and \mbox{FWHM = 2.72
  pixels}, in the same order. All values were determined averaging the
FWHM of several stars inside the field of view, selected
because they showed no saturation and were located close to the
blazars (see all stars labeled in Figure~\ref{fig:fov}).

To represent the surface brightness profile of the host galaxy we used
Sersic's law:
\begin{equation}
I(a_{\mathrm{HG}})=I_{\mathrm{o}}\times\mathrm{e}^{-(a_{\mathrm{HG}}/r_{\mathrm{0}})^{1/n}},
\label{eq:sersic}
\end{equation}
\noindent where $I_{\mathrm{o}}$ is the central surface brightness,
$r_{\mathrm{0}}$ is a pseudo scale parameter, and $n$ ($n>0$)
corresponds to the Sersic's index, a parameter that determines the
shape of the brightness profile. All parameters were fitted to the
data. Once the best fit parameters were obtained, we determined the
structure, the flux, and the instrumental magnitude of the host
galaxy. All fit parameters are summarized in the upper part of
Table~\ref{tab:params_gal}.

\subsection{Recovery of the intrinsic parameters of the host galaxies}

To characterize the effect of the FWHM on the previously determined
structural parameters, instead of following the approach described in
\citet{Trujillo2001}, we build up an empirical relationship, with the
goal of recovering the intrinsic parameters of the host galaxy when
the observed ones are used as input. Although the latter is not a
direct method, we opted for this more conservative treatment. The
sampling of our data is not optimal for 2-D fitting algorithms such as
GALFIT \citep{peng2002} to give reliable results. This was concluded
after several trials on simulated images with similar characteristics
as our observations. We finally opted for an approach which uses
synthetic data, allowing us to better understand the behavior and
correlations between parameters when seeing changes, thus identifying
regions of parameter space where the recovery of intrinsic parameters
might not be reliable. To this end we generated simulated host
galaxies using Sersic's law (Eq. \ref{eq:sersic}) changing the values
for $a_{HG}$ and $n$, and arbitrarily fixing $I_{\mathrm{o}}$. These
images were then convolved with a Gaussian function with different
values of FWHM to account for different values of seeing. We obtained
1890 images, that we re-analyzed to recover their structural
parameters in the exact same fashion as discussed in
Sect.~\ref{sec:params_struc}. Analyzing the parameters obtained by the
fitting algorithms, and comparing them to the ones used to generate
the synthetic data, we found that they are not directly related in a
one-by-one fashion, but following some relations that involve other
structural parameters. For example, we found that the fitted effective
radius depends on the effective radius used to create the synthetic
image, but also on the Sersic's index. This is nothing more than the
reflection of a strong coupling between parameters. Using these as
empirical relations we input our previously determined structural
parameters, and obtained the intrinsic (i.e., seeing free)
parameters. The lower panel of Table~\ref{tab:params_gal} summarizes
our results. After a quick comparison between the top (observed) and
bottom (recovered) parameters of the two host galaxies we see, for
instance, that the observed Sersic's index is systematically smaller
than the recovered one. A lower $n$ would imply a less steep
brightness profile, which is exactly the impact that seeing has over
point sources \citep{Trujillo2001}.

\begin{table}[ht!]
  \caption{Parameters of model fits for both blazars in the $R$ and
    $B$ bands. {\it Top:} parameters derived from the
    observations. {\it Bottom:} recovered intrinsic parameters of the
    host galaxies once seeing has been accounted for.}
  \label{tab:params_gal}
  \centering 
  \begin{tabular}{c c c c c} 
    \hline\hline
    Band  &   $a_{\mathrm{HG}}$      &    $n$    &    I$_0$      &    m$_{\mathrm{gal}}$   \\
          &   (pix)        &           & (ADU/pix$^2$) &     (mag)     \\
    \hline
    \multicolumn{5}{c}{\mbox{1ES 1959+650, observed}}\\
    $B$     &   14.3564      &  1.9523   &  486.31       &  16.295       \\
    $R$     &   12.5675      &  1.7838   & 1941.50       &  14.033       \\
    \multicolumn{5}{c}{\mbox{HB89 2201+044, observed}}\\
   $B$     &   14.3568      &  2.1363   &  1234.52      &  15.391       \\
    $R$     &   18.7721      &  2.1779   & 7471.17       &  12.875       \\
    \hline
    \multicolumn{5}{c}{\mbox{1ES 1959+650, recovered}}\\
    $B$     &   14.5169      &  1.9734   &  476.85       &  16.332       \\
    $R$     &   12.5264      &  1.8156   & 1872.09       &  14.140       \\
    \multicolumn{5}{c}{\mbox{HB89 2201+044, recovered}}\\    
    $B$     &   14.5017      &  2.1615   & 1221.48       &  15.429       \\
    $R$     &   20.0153      &  2.2351   & 8373.34       &  12.723       \\
    \hline\hline
  \end{tabular}
\end{table}

\subsection{Correction of the polarimetric measurements by the contribution of the host galaxy}

To correct for the depolarizing effect that the host galaxy introduces
in our measurements, we have to take into account the following
relation:
\begin{equation}
P = P_{\mathrm{obs}}\left(1 - \frac{F_{\mathrm{G}}}{F_{\mathrm{AGN}} + F_{\mathrm{G}}}\right)^{-1}\;,
\label{eq:polcorr}
\end{equation}
\noindent where $F_{AGN} + F_G$ is the observed standard flux of the
AGN plus host galaxy, and $F_G$ is the standard flux of the host
galaxy, both wavelength dependent and at this point unknown
\citep[see][for a full description on the formulas of
  Eq.~\ref{eq:polcorr}]{ACR08}. To estimate $F_G$ we made use, one
more time, of synthetic data. We simulated a host galaxy per band,
using as input parameters the intrinsic parameters listed in the
bottom panel of Table~\ref{tab:params_gal}, and integrated the flux
inside the aperture that was originally considered to extract the
fluxes of real data (see Sect.~\ref{sec:OaDR} for further
details). However, before this we convolved this synthetic image with
a Gaussian kernel, whose standard deviation reflected the seeing of
the true data, not globally but image by image. In particular, this
seeing was estimated averaging the seeing values of unsaturated stars
in the field of view. As an example, if during a given night we
collected four images, each one with one of the four position angles
of the rotating plate, we calculated the average seeing of each image
and used it to convolve the image of the host galaxy. We end up with
four synthetic images, that were integrated to determine the expected
standard flux of the host galaxy affected by the time-dependent
seeing. However, we still need to overcome the fact that $F_{AGN} +
F_G$ is still unknown. To estimate this we can assume the following
relation:
\begin{equation}
\frac{f_*}{F_*} = \frac{f_{\mathrm{AGN}} + f_{\mathrm{G}}}{F_{\mathrm{AGN}} + F_{\mathrm{G}}}\;,
\end{equation}
\noindent where, neglecting color-dependent terms, the relation shows
that the flux ratio between instrumental and standard fluxes of a
given reference star, $\frac{f_*}{F_*}$, should equal the flux ratio
between the instrumental and standard fluxes of the blazar plus host
galaxy, $\frac{f_{\mathrm{AGN}} + f_G}{F_{\mathrm{AGN}} + F_G}$.

To calculate the standard magnitudes (and thus fluxes, $F_*$) of the
reference stars, we used data of photometric standard stars of two
Landolt fields, \mbox{SA $115$} and \mbox{SA $114$}
\citep{Landolt1992}, that we acquired during a photometric night and
without polarizer. For the stars in the field of \mbox{1ES 1959+650},
our results agree with those by \citet{Pace2013}. We found no
previously published standard magnitudes for stars in the field of
HB89 2201+044, so we report them in Appendix \ref{sec:appendix}.

\begin{figure}[ht!]
  \centering
  \includegraphics[width=.5\textwidth]{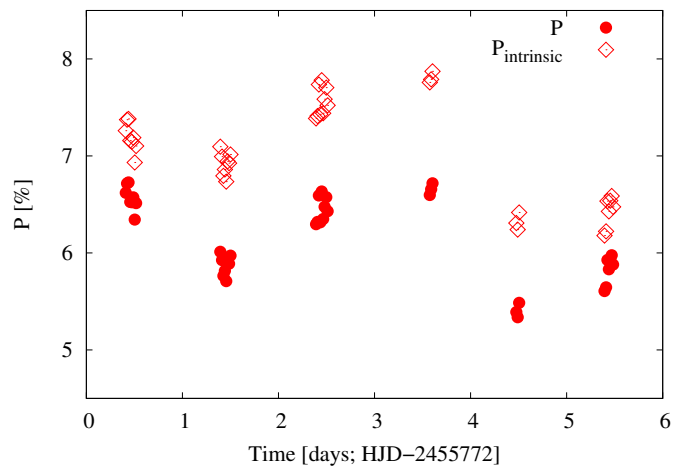}
  \caption{\label{fig:pol_GI} Polarimetric behavior of the blazar
    \mbox{1ES 1959+650} along the whole campaign in the $R$ band,
    corrected only by instrumental polarization (P, red filled
    circles) and instrumental polarization plus the contribution of
    the host galaxy (P$_{\mathrm{intrinsic}}$, red empty diamonds).}
\end{figure}

With $F_*$ computed we obtained, image by image, \mbox{$F_{AGN} +
  F_G$}. Using this quantity in Eq.~\ref{eq:polcorr} we calculated
four correction terms per polarimetric point (again, each one
corresponding to each rotation angle, and each one with its
respective seeing value) and calculated a last correction term by
averaging these four. Finally, we re-computed the polarization degree
along the whole campaign taking into account this correction. The
results of polarization degree for the whole campaign, for the $R$
band and the blazar \mbox{1ES 1959+650}, can be seen in
Figure~\ref{fig:pol_GI}. In this case, the averaged difference between
both polarization states is of the order of 1\% (see
Table~\ref{tab:pol_GI} for a complete picture of both blazars and both
bands).

Furthermore, as described in Sect.~\ref{sec:polarimetry}, we
re-calculated the probability values associated to intra and
inter-night variability, but now using the polarimetric points
corrected by the depolarizing effect introduced by the host
galaxies. Although results do not change (we found again no
intra-night variability but inter-night variability) the probability
values significantly decrease, around a factor of 3, which would
point out the need of correcting for the effect described in this
Sect.

\begin{table}[ht!]
  \caption{Mean values for the polarization degree with ($P$) and without
    ($P_{\mathrm{intrinsic}}$) the contribution of the host galaxy for
    the blazars \mbox{1ES 1959+650} and \mbox{HB89 2201+044}.}
  \centering
  \begin{tabular}{c c c c}
    \hline \hline
    Blazar                 &    $\langle P \rangle$         &  $ \langle P_{\mathrm{intrinsic}} \rangle $   & Band  \\
    \hline
    \mbox{1ES 1959+650}  & 6.97 $\pm$ 0.50 & 7.64 $\pm$ 0.64 & $B$ \\
    \mbox{1ES 1959+650}  & 6.17 $\pm$ 0.41 & 7.06 $\pm$ 0.49 & $R$ \\
    \mbox{HB89 2201+044} & 0.70 $\pm$ 0.46 & 1.01 $\pm$ 0.67 & $B$ \\
    \mbox{HB89 2201+044} & 0.38 $\pm$ 0.30 & 0.62 $\pm$ 0.49 & $R$ \\
    \hline
  \end{tabular}
  \label{tab:pol_GI}
\end{table}

\subsection{Impact of seeing on our polarimetric measurements}

\begin{figure*}[ht!]
  \centering
  \includegraphics[width=.45\textwidth]{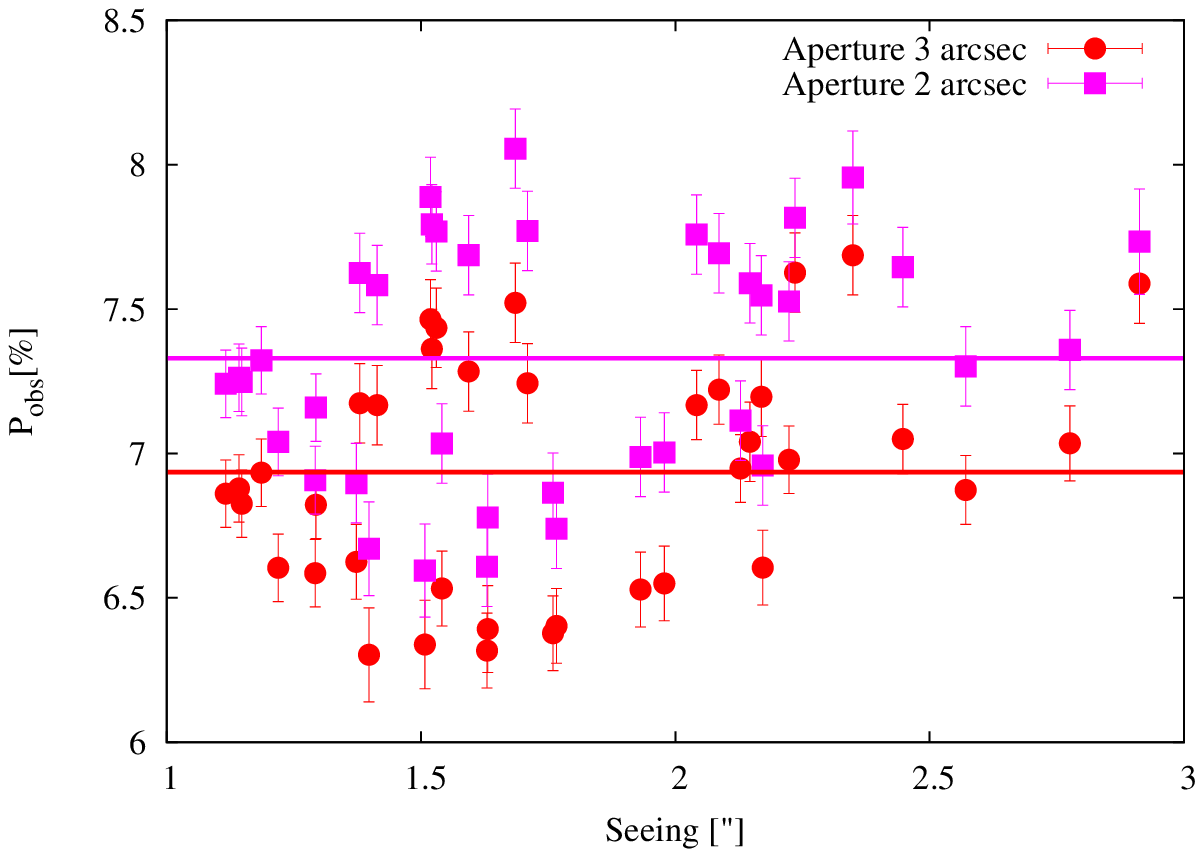}
  \includegraphics[width=.45\textwidth]{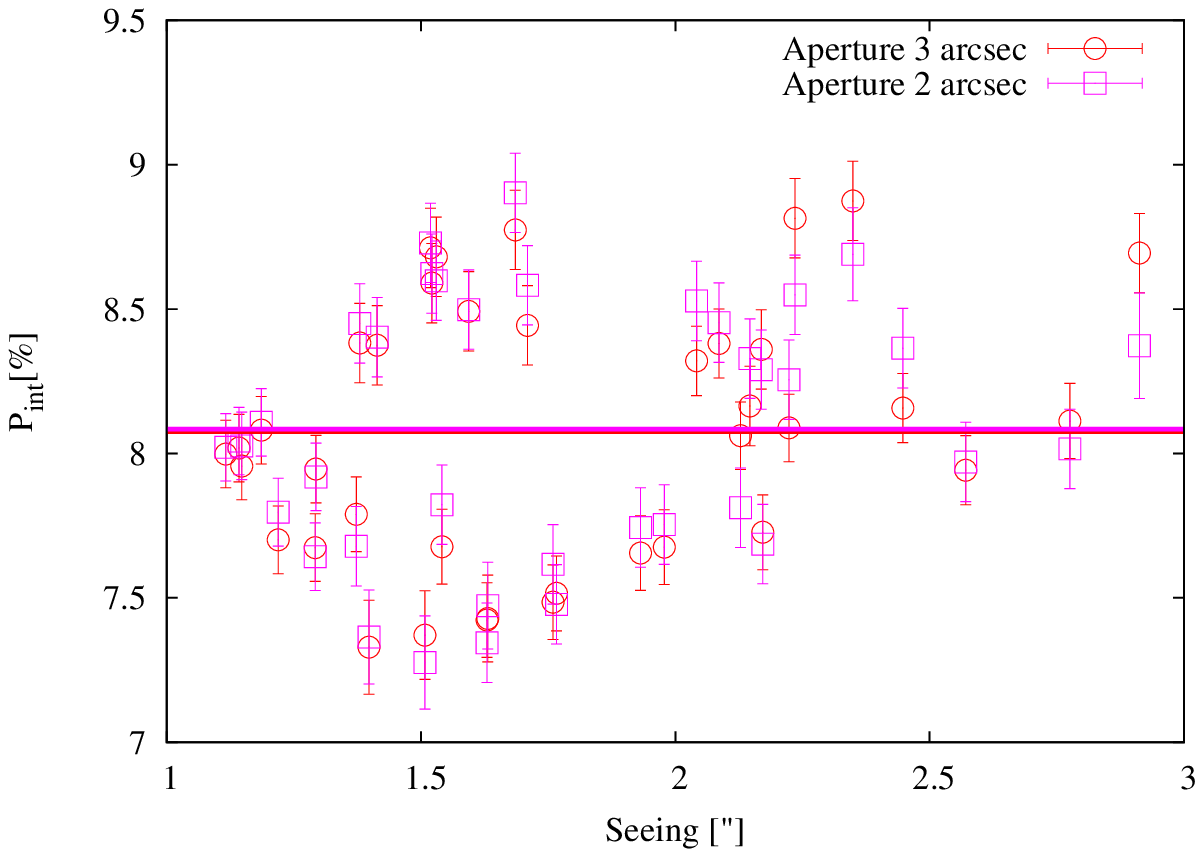}
  \caption{\label{fig:seeing_pol} Behavior of the polarization degree
    with seeing for \mbox{1ES 1959+650} in the $R$ band when two
    fixed apertures are considered. In the left panel data points
    correspond to the observed polarization, $P_{\mathrm{obs}}$, while
    the right panel shows the polarization corrected by the
    contribution of the host galaxy,
    $P_{\mathrm{intrinsic}}$. Horizontal lines following the same
    color-code indicate mean values of polarization and are plotted to
    guide the readers eye.}
\end{figure*}

As previously mentioned, the changes in the photometric quality during
the observing nights may introduce spurious variability that has to be
considered when microvariability is being reviewed. Seeing strongly
degrades measurements. Furthermore, as a result of seeing variability
a given fixed aperture introduces a larger or lower amount of
unpolarized light coming from the host galaxy. As a result, the
depolarizing effect changes with seeing, affecting the measurements of
polarized light of the nucleus, if not corrected. Seeing variability
directly reflects into the flux ratio between the nucleus and the
galaxy, because both components have a different brightness
distribution and are, in consequence, differently affected. Our seeing
values range between 1 and \mbox{3 arcsec}, as seen on the horizontal
axis of Figure~\ref{fig:seeing_pol}. As mentioned in previous sections,
our chosen aperture was set to be \mbox{3 arcsec} and was considered
fixed along the whole campaign. Therefore, the chosen aperture is
larger than the seeing values along the campaign and always contains,
as consequence, most of the flux of the nucleus \citep{H89}. To test
the impact of seeing in our polarimetric measurements, and the
importance of correcting for the depolarizing effect of the host
galaxy, we did the following exercise. First, we re-calculated all our
polarimetric (and thus photometric) quantities using an aperture of
\mbox{2 arcsec}. Our results for the $R$ band and the blazar \mbox{1ES
  1959+650} can be seen in the left part of
Figure~\ref{fig:seeing_pol}. While the red filled circles show the
polarimetric measurements for an aperture equal to \mbox{3 arcsec},
the pink filled squares show polarimetric values for an aperture of
\mbox{2 arcsec}. Neither of them have been corrected for the
depolarizing effect introduced by the host galaxy. As the Figure
clearly shows, the polarization is larger for the smaller
aperture. This is simply because a smaller aperture implies a smaller
contribution of the host galaxy depolarizing effect which, in turn,
translates into a larger polarization. Then, we corrected for the
depolarizing contribution of the host galaxy for both apertures, as
explained in previous Sections. Our results can be seen in the right
panel of Figure~\ref{fig:seeing_pol}, following the same color and
point-code. Not surprisingly, the polarization level of both apertures
is on average the same (and higher than uncorrected values), showing
again the relevance of correcting for the host galaxy.

\subsection{Corrected photometry along the campaign}

\begin{figure}[ht!]
  \centering
  \includegraphics[width=.5\textwidth]{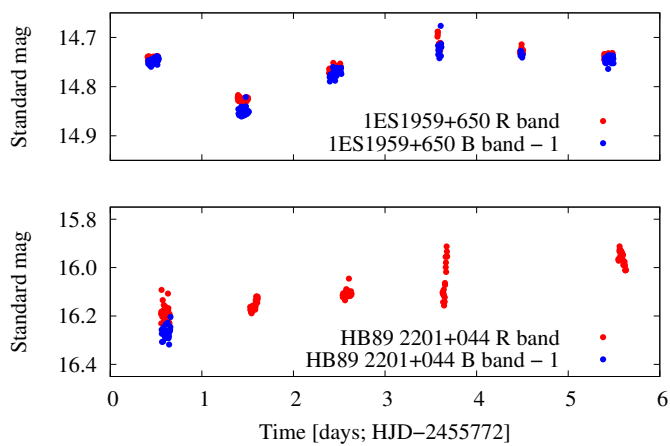}
  \caption{\label{fig:standard_mags} Standard magnitudes of the blazar
    without the contribution of the host galaxy for \mbox{1ES
      1959+650} ({\it top}) and \mbox{HB89 2201+044} ({\it
      bottom}) for both filters ($R$-band in red, $B$-band in
    blue). $B$-band values were shifted in 1 magnitude for a better
    visualization.}
\end{figure}

Figure~\ref{fig:standard_mags} shows the photometric behavior of the
two blazars along the whole campaign, when the contribution of the
host galaxy has been removed. For a better visualization, both $B$
band quantities were shifted by one magnitude. We observed photometric
variability in both blazars along the campaign, and a similar trend in
inter-night variability for the case of \mbox{1ES 1959+650}. We
observed no significant intra-night variability. In particular, for
\mbox{1ES 1959+650}, our results are similar to the values found by
\cite{Sorcia2013intro3}, where the source showed a minimum and maximum
of brightness of $R=15.2$ mag and $R=14.08$ mag, respectively.

\section{Discussion and conclusions}
\label{sec:Discussion}

Blazars are known to have extreme photo-polarimetric variability. Some
examples of these are \mbox{AO 0235+164} \citep{Cellone2007}, or
\mbox{PKS 1510-089} \citep{Aleksic2014}. The last object reached a
peak flux of $18$ mJy in the $R$ band while the quiescent level flux
is typically $\sim 2$ mJy.  During this major optical flare, the
optical polarization degree increased to $> 30 \%$. This reveals the
importance of photo-polarimetric follow-ups of blazars, with the main
goal to understand and properly model the source of their
variability. In this work we have undertaken a photo-polarimetric
follow-up that included two targets, an HBL object \mbox{(1ES
  1959+650)} and an LBL object \mbox{(HB89 2201+044)}. During our
observations, we simultaneously registered their polarimetric and
photometric behavior. We analyzed the behavior of their linear
polarization computing the parameters $P$ and $\Theta$ throughout the
whole campaign. \mbox{1ES 1959+650} seems to present a very moderate
inter-night variability, while the angle has a slight ($\sim 10$~deg.)
rotation along the campaign. As we mentioned in Sect.~\ref{sec:OaDR},
the HBLs should show statistically lesser amounts of polarized light
than that of LBLs. This behavior is not in agreement with our results,
since we found a $P \sim 7 \%$ for \mbox{1ES 1959+650}, while
\mbox{HB89 2201+044} shows a polarization consistent with zero when
errors at two-sigma level are considered ($P \sim 0.70\%$). Regarding
\mbox{HB89 2201+044}, one explanation could be that the object is
currently in a low activity state. There are not enough data nor
literature available to say the contrary. Further simultaneous
photometric and polarimetric observations of this object are
required. Carrying out a statistical analysis, we find a non-detection
of intra-night polarimetric microvariability in both blazars, while a
significant polarimetric variability is evident when the whole
campaign is taken into consideration. We also observe a very moderate
inter-night photometric variability for \mbox{HB89 2201+044} in filter
$B$, and in both filters for \mbox{1ES 1959+650}.

Our targets are relatively nearby objects. Their host galaxies
introduce a depolarizing effect which, in turn, can lead to systematic
errors in the derived photo-polarimetric quantities when seeing
conditions vary with time. We have modeled the incidence of the host
galaxy and, for the first time, we have corrected our polarimetric
data by the depolarizing effect introduced by the host galaxy in an
auto-consistent way, this is, using our own data to obtain its
structural parameters. Simultaneously, we have considered spurious
variability introduced by varying seeing conditions into our
microvariability analysis. Comparing our values with and without
correcting for the host galaxies, the intrinsic polarization is $1\%$
and $0.3\%$ higher for \mbox{1ES 1959+650} and \mbox{HB89 2201+044},
respectively, while the behavior of intrinsic polarization with time
is the same than the observed polarization in both blazars. For the
case of \mbox{1ES 1959+650}, if we compute the ratio of polarizations
in $B$ and $R$ bands, not taking into consideration the host galaxy,
gives $P_{\mathrm{B}}/P_{\mathrm{R}}= 1.12 \pm 0.02$. This value can
be explained as follows: blazars are in elliptical galaxies which
present dominant starlight emission in $R$ band. In consequence, the
depolarizing effect introduced by the host galaxy is smaller in the
$B$ band than in the $R$ band. Therefore, the observed polarization in
$B$ is expected to be larger than in $R$. After applying the host
galaxy correction, we obtained $P_{\mathrm{B}}/P_{\mathrm{R}} =1.08
\pm 0.02$, which indicates that the polarization is almost the same in
both band. The difference between both ratios is significant at a
$\sim 1.5 \,\sigma$ level. Larger differences could be obtained under
different atmospheric conditions and for blazars with other
host-galaxy to AGN flux properties. Therefore, if we do not take into
account the effect introduced by the host galaxy there would be a
tendency to retrieve erroneous results. This could be relevant in
studies of frequency dependent polarization \citep[see
  e.g.,][]{BarresdAlmeida2010}. Finally, the presence of dust features
in host galaxies, such as the case of \mbox{1ES 1959+650}
\citep[][]{Heidt1999}, may be another source of uncertain. Their
effects depend on wavelength, so they could affect polarization
measurements in a different amount in each photometric band. However,
we believe the quality of our data is not sufficient to recognize this
effect from our polarimetric uncertainties.

In general, our work shows that if the host galaxy is not properly
taken into account, and also if changing seeing conditions are not
take care of, a significant error in the computation of the
polarization degree of blazars can be produced. This, in turn, could
end up in misleading models or conclusions derived from erroneous
polarization states. And the spurious results are intensified if we
study high polarized objects, such as the HBL type.

%__________________________________________________________________

\begin{acknowledgements}
     The authors wish to acknowledge the referee's comments. We are
     deeply grateful to Santos Pedr\'az, Jes\'us Aceituno, and Calar
     Alto staff for their invaluable help during the
     observations. This work was funded with grants from Consejo
     Nacional de Investigaciones Científicas y Técnicas de la
     República Argentina and Universidad Nacional de La Plata
     (Argentina). Funding for the Stellar Astrophysics Centre is
     provided by The Danish National Research Foundation (grant
     No. DNRF106).
\end{acknowledgements}

\bibliographystyle{aa} % style aa.bst
\bibliography{AA-Sosa2017} % your references Yourfile.bib

%\end{document}
%

%

%-------------------------------------------------------------
%               Appendices have to be placed at the end, after
%                                        \end{thebibliography}
%-------------------------------------------------------------

\begin{appendix}
\section{Standard magnitudes}
\label{sec:appendix}
\begin{figure}[ht!]
  \centering
  \includegraphics[width=.45\textwidth]{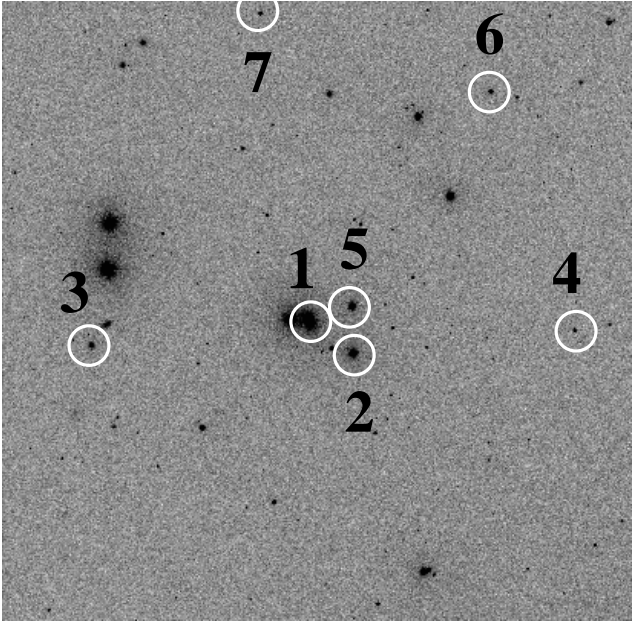}
  \caption{\label{fig:2201_SinP}The field of view without polarizer
    ($9$ $\times$ $9$ arcmin) of \mbox{HB89 2201+044}. The white
    circles indicate the locations of the blazar (1) and field stars
    ($>$1). East is up and North is to the right.}
\end{figure}

Standard magnitudes and associated 1$\sigma$ errors of the field stars
of \mbox{HB89 2201+044} can be found in
Table~\ref{tab:standardMag}. These standard magnitudes were obtained
following the procedure described in Sect.~\ref{sec:results2}. The
labels of Table~\ref{tab:standardMag} correspond to those of
Figure~\ref{fig:2201_SinP}. For stars in the field of \mbox{1ES
  1959+650}, the standard magnitudes and their errors are in agreement
with \citet{Pace2013}.

\begin{table}[ht!]
  \caption{Standard magnitudes of the field stars of \mbox{HB89
      2201+044} in $R$ and $B$ bands.}  \centering
  \begin{tabular}{c c c}
    \hline \hline
        Star      &    $R$  & $B$  \\
    \hline
                              2 & 13.427$\pm$0.004   & 14.190$\pm$0.002 \\
                              3 & 14.423$\pm$0.007   & 15.549$\pm$0.004 \\
                              4 & 15.527$\pm$0.022   & 17.671$\pm$0.011 \\
                              5 & 13.116$\pm$0.004   & 14.678$\pm$0.002 \\
                              6 & 14.159$\pm$0.014   & 16.793$\pm$0.007 \\
                              7 & 15.718$\pm$0.020   & 17.337$\pm$0.010  \\

    \hline
    \end{tabular}
  \label{tab:standardMag}
\end{table}
\end{appendix}

\end{document}